\documentclass[letter]{ieice}
\usepackage[dvips]{graphicx}
\field{}
\title{ Bounded-Choice Statements for  User Interaction  in
 Imperative  Programming}
\authorlist{
\authorentry{Keehang KWON}{m}{labelA}
\authorentry{Jeongyoon SEO}{n}{labelA}
\authorentry{Daeseong Kang}{n}{labelB}
}
\affiliate[labelA]{The authors are with Computer Eng., DongA University. email:khkwon@dau.ac.kr}
\affiliate[labelB]{The author is with Electronics Eng., DongA University.}
\received{2003}{1}{1}
\revised{2003}{1}{1}
\finalreceived{2003}{1}{1}

\makeatletter
\long\def\@makemyfntext#1{$^{\rm *}\ $ #1}

\long\def\@myfootnotetext#1{\insert\footins{\footnotesize
    \interlinepenalty\interfootnotelinepenalty 
    \splittopskip\footnotesep
    \splitmaxdepth \dp\strutbox \floatingpenalty \@MM
    \hsize\columnwidth \@parboxrestore
   \edef\@currentlabel{\csname p@footnote\endcsname\@thefnmark}\@makemyfntext
    {\rule{\z@}{\footnotesep}\ignorespaces
      #1\strut}}}

\def\myfootnotetext{\@ifnextchar
     [{\@xfootnotenext}{\xdef\@thefnmark{\thempfn}\@myfootnotetext}}
\makeatother

\newenvironment{numberedlist}
{\begin{list}{\makebox[20pt]{\hss(\arabic{itemno})\enspace}}
             {\usecounter{itemno}\labelwidth 20pt}}{\end{list}}

\newcounter{itemno}

\newcounter{itemno1}

\newcounter{itemno2}

\newcounter{exno}

\newcounter{defno}

\newenvironment{defn}{\refstepcounter{defno}\medskip \noindent {\bf
Definition \thedefno.\ }}{\medskip}

\newcommand{\sep}{\;\vert\;}

\newcommand{\oprove}{\vdash\kern-.6em\lower.7ex\hbox{$\scriptstyle O$}\,}

\newcommand{\Pscr}{{\cal P}}

\newcommand{\pderivation}{{\cal P}\kern -.1em\hbox{\rm -derivation}}
\newcommand{\pderivationl}{{\cal P}\kern -.1em\hbox{\em -derivation}}
\newcommand{\pderivable}{{\cal P}\kern -.1em\hbox{\rm -derivable}}
\newcommand{\pderivablel}{{\cal P}\kern -.1em\hbox{\em -derivable}}
\newcommand{\pderivations}{{\cal P}\kern -.1em\hbox{\rm -derivations}}
\newcommand{\pderivability}{{\cal P}\kern -.1em\hbox{\rm -derivability}}

\newcommand{\all}{\forall}

\newcommand{\ie}{{\em i.e.}}

\newsavebox{\lpartfig}
\newsavebox{\rpartfig}

\newenvironment{exmple}{
 \begingroup \begin{tabbing} \hspace{2em}\= \hspace{3em}\= \hspace{3em}\=
\hspace{3em}\= \hspace{3em}\= \hspace{3em}\= \kill}{
 \end{tabbing}\endgroup}

\newcommand{\lb}{\langle}
\newcommand{\rb}{\rangle}
\newcommand{\pr}{prov}

     {\\* \hspace*{\fill} \end{trivlist}}

\newcommand{\muprolog}{{C$^{BI}$}}
\newcommand{\uch}{kchoose}
\newcommand{\kch}{kchoose}
\renewcommand{\pr}{ex}

\begin{document}
\maketitle
\begin{summary}
 Adding versatile interactions to imperative  programming -- C, Java and
Android -- is an essential
task. 
Unfortunately, existing  languages provide only limited constructs for user interaction.
These constructs are usually in the form of $unbounded$ quantification. For example,
  existing  languages can take the keyboard input from the user only
via the $read(x)/scan(x)$  statement. Note that the value of $x$  is  unbounded
in the sense that $x$ can have any value. This statement is thus not useful for 
applications with bounded
inputs.
  To support bounded choices, we propose new bounded-choice statements for user interation.
Each input device (keyboard,  mouse, touchpad, $\ldots$) 
naturally requires a new  bounded-choice statement.
To make things simple, however, we focus on a  bounded-choice statement  for keyboard -- $\kch$ --
 to allow for more controlled and more 
guided participation from the user. We illustrate our idea
via \muprolog, an extension of the core C  with a new 
bounded-choice statement for the keyboard.
\end{summary}

\section{Introduction}\label{sec:intro}

 Adding versatile interactions to imperative  programming -- C, Java,
Android, etc. -- has become an essential
task. 
Unfortunately, existing  languages provide only limited constructs for user interaction.
These constructs are usually in the form of $unbounded$ quantification. For instance, 
the keyboard input statement  that has been used
in Java-like languages  is restricted to the $read/scan$ statement. 
The $read$ statement is  of the form $read(x); G$, where $G$ is a statement and 
$x$ can have any value. Hence, it is a form of an $unbounded$ quantified 
statement.  However, in many situations,
the system requires a  form of  
{\it bounded-choice} interactions; the user is expected to choose  one among many alternatives.  Examples include most interactive systems such as airline ticketing systems.

The use of bounded-choice interactions is thus essential  in 
representing most interactive
systems. For this purpose, this paper proposes  a bounded-choice approach to user 
 interaction. Each input device naturally requires a new
bounded-choice statement.   To make things simple,  however, 
we focus only on the keyboard device. It is straightforward
to adjust our idea to other input devices such as  mouse and  touchpad.

Toward this end, we propose a new bounded keyboard input statement 
$\kch(G_1,\ldots,G_n)$, 
where each  $G_i$ is a statement. This has the following execution semantics:

\[ \pr(\Pscr, \uch(G_1,\ldots,G_n))\ \leftarrow\  \pr(\Pscr,  G_i) \] 

\noindent , where $i$ is  chosen (\ie, a keyboard input) by the user and $\Pscr$ is a 
set of procedure definitions.  The notation $S \leftarrow R$ denotes the reverse implication, \ie, 
$R \rightarrow S$.
In the above definition, the system  requests the user to choose $i$ via the keyboard 
and then proceeds
with executing $G_i$.  If $i$ is not among $\{ 1,\ldots,n \}$, then we assume that the system does 
nothing.
It can be easily seen that our new statement has many applications
 in representing most interactive systems.

The following C-like code example reads a variable named $emp$ from the keyboard,
 whose value represents an employee's name. 

\begin{exmple}
        read(emp); \\
        switch (emp) \{ \\
 \>           case tom:  age = 31;   break; \\
  \>          case kim: age = 40;   break;\\
 \>           case sue: age = 22;    break;\\
 \>           default: age = 0;          \\
        \}\\
\end{exmple}

\noindent 
In the above, the system 
 requests the user to type in a particular employee. 
 Note that the above code is based on unbounded quantification
 and is thus very awkward. It is also  error-prone 
 because the user may type in  an invalid value.

The above application obviously
 requires a bounded-choice interaction rather than  one based on
unbounded quantification.
Our $\uch$ statement provides such a  bounded-choice interaction for  keyboard   
 and is useful to avoid this kind of human error.
Hence, instead of the above code, consider the statement

\begin{exmple}
print(``Enter 1 for tom, 2 for kim and 3 for sue:''); \\
        \kch( \\
 \>           emp = tom;  age = 31, \\
  \>          emp = kim; age = 40, \\
 \>           emp = sue; age = 22 );
\end{exmple}

\noindent This program expresses the task of the user choosing one among
three employees.
 Note that this program is much easier and safer to use.
 The system now requests the user to select one (by typing 1,2,3)  
among three
employees. After it is selected, 
the system sets his age as well.

Generally speaking, the $\kch$ statement is designed to directly encode most interactive objects
which require  the user to choose one among several possible tasks.
Hence, there is a rich realm of applications for this statement.
For example, as we will see later in Section 3, the ATM machine requires the user
to select one among 1) balance checking, 2) cash withdrawal, and 3) cash deposit.
Hence, it can be directly encoded via the $\kch$ statement.

 It is easy to observe that  $\kch$ statement  can be built from the
 $read$-$switch$ combination. For example, the above example can be rewritten in the following way.

\begin{exmple}
print(``Enter 1 for tom, 2 for kim and 3 for sue:''); \\
        read(n); \\
        switch (n) \{ \\
 \>           case 1:  emp = tom; age = 31;   break; \\
  \>          case 2:  emp = kim; age = 40;   break;\\
 \>           case 3:  emp = sue; age = 22;    break;\\
 \>           default:           \\
        \}\\
\end{exmple}


 It is then tempting to conclude that the $\kch$ construct is not needed because it can be built from the
 $read$-$switch$
combination. However, this view is quite misleading. Without it, the resulting codes would be 
low-level for the following reasons: 

\begin{itemize}

\item The programmer must manually allocate a variable for the
$read$ construct.

\item The programmer must specify the numbering sequence in the $switch$ statement.

\item The programmer must specify the default part.

\end{itemize}
\noindent
As a consequence, these codes are cumbersome, error-prone, difficult to read,  and reason about.

 The $\kch$  construct should rather be seen as a well-designed, 
high-level abstraction for bounded-choice interaction and the  
$read$-$switch$ combination should  be seen as  its low-level implementation. 
 The advantage of the use of this construct becomes evident when an application has a long
sequence of  interactions with the user. Therefore, the need for this construct is clear.
 To our knowledge, this kind of construct  has never been proposed before in imperative languages.
This is quite surprising, given the ubiquity of
  bounded-choice interaction in interactive applications.

The $\kch$  construct can be implemented in many ways. One way to implement the $\kch$  construct is via preprocessing, \ie, via transformation to plain C-like code.
That is,  $\kch(G_1,\ldots,G_n)$ is transformed to the following: 

\begin{exmple}
         int k; \\
        read(k); \\
        switch (k) \{ \\
 \>           case 1:  $G'_1$;   break; \\
  \>          case 2:  $G'_2$;   break;\\
\> \vdots \\
 \>           case n:  $G'_n$;    break;\\
 \>           default:           \\
        \}\\
\end{exmple}
\noindent Here, k is a new, local storage, and $G'_1,\ldots,G'_n$ are obtained from
 $G_1,\ldots,G_n$ via the same transformation.

This paper focuses on the minimum 
core of C. This is to present the idea as concisely as possible.
The remainder of this paper is structured as follows. We describe 
 \muprolog, an extension of  core C  with a new 
bounded-choice statement for the keyboard
 in Section 2. In Section \ref{sec:modules}, we
present an example of  \muprolog.
Section~\ref{sec:conc} concludes the paper.

\section{The Language}\label{sec:logic}

The language is   core C 
 with  procedure definitions. It is described
by $G$- and $D$-formulas given by the syntax rules below:
\begin{exmple}
\>$G ::=$ \>   $true \sep A \sep x = E \sep  G;G \sep  read(x);G  \sep $ \\  
\>\>  $\kch(G_1,\ldots,G_n)$ \\
\>$D ::=$ \>  $ A = G\ \sep \all x\ D$\\
\end{exmple}
\noindent
 In the above, 
$A$ in $D$ represents a head of an atomic procedure definition of the form $p(x_1,\ldots,x_n)$ 
where $x_1,\ldots,x_n$ are parameters. $A$ in $G$ represents a procedure call
 of the form $p(t_1,\ldots,t_n)$ 
where $t_1,\ldots,t_n$ are actual arguments.
A $D$-formula  is called a  procedure definition.
In the transition system to be considered, $G$-formulas will function as the
main  statement, and a set of $D$-formulas  enhanced with the
machine state (a set of variable-value bindings) will constitute  a program.
Thus, a program is a union of two disjoint sets, \ie, $\{ D_1,\ldots,D_n \} \cup \theta$
where each $D_i$ is a $D$-formula and $\theta$ represents the machine state.
Note that $\theta$ is initially set to an empty set and will be updated dynamically during execution
via the assignment statements. 

 We will  present an interpreter via a proof theory \cite{Khan87,MNPS91,HM94,MN12}.
Note that this interpreter  alternates between 
 the execution phase 
and the backchaining phase.  
In  the execution phase (denoted by $ex(\Pscr,G,\Pscr')$) it tries to execute a main statement $G$ with respect to
a program $\Pscr$ and
produce a new program $\Pscr'$
by reducing $G$ 
to simpler forms until $G$ becomes an assignment statement or a procedure call. The rules
 (6), (7),(8) and (9) deal with this phase.
If $G$ becomes a procedure call, the interpreter switches to the backchaining mode. This is encoded in the rule (3). 
In the backchaining mode (denoted by $bc(D,\Pscr,A,\Pscr')$), the interpreter tries 
to solve a procedure call  $A$ and produce a new  program $\Pscr'$
by first reducing a procedure definition $D$ in a program $\Pscr$ to  its instance
 (via rule (2)) and then backchaining on the resulting 
definition (via rule (1)).
 To be specific, the rule (2) basically deals with argument passing: it eliminates the universal quantifier $x$ in $\all x D$
by picking a value $t$ for
$x$ so that the resulting instantiation, written as $[t/x]D$, matches the procedure call $A$.
 The notation $S$\ seqand\ $R$ denotes the  sequential execution of two tasks. To be precise, it denotes
the following: execute $S$ and execute
$R$ sequentially. It is considered a success if both executions succeed.
Similarly, the notation $S$\ parand\ $R$ denotes the  parallel execution of two tasks. To be precise, it denotes
the following: execute $S$ and execute
$R$  in any order.  Thus, the execution order is not important here. 
It is considered a success if both executions succeed.
 The notation $S$\ choose\ $R$ denotes the  selection between two tasks. To be precise, it denotes
the following: the machine selects and executes  one between  $S$ and\ $R$.
 It is considered a success if the selected one succeeds.

As mentioned in Section 1, the notation $S \leftarrow R$ denotes  reverse implication, \ie, $R \rightarrow S$.

\begin{defn}\label{def:semantics}
Let $G$ be a main statement and let $\Pscr$ be a program.
Then the notion of   executing $\lb \Pscr,G\rb$ successfully and producing a new
program $\Pscr'$-- $ex(\Pscr,G,\Pscr')$ --
 is defined as follows:
\begin{numberedlist}

\item    $bc((A = G_1),\Pscr,A,\Pscr_1)\ \leftarrow$ 
 $ex(\Pscr, G_1,\Pscr_1)$. \% A matching procedure for $A$ is found.

\item    $bc(\all x D,\Pscr,A,\Pscr_1)\ \leftarrow$    $bc([t/x]D,
\Pscr, A,\Pscr_1)$. \% argument passing

\item    $ex(\Pscr,A,\Pscr_1)\ \leftarrow$    $(D \in \Pscr$ parand $bc(D,\Pscr, A,\Pscr_1))$. \% a procedure call

\item  $ex(\Pscr,true,\Pscr)$. \% True is always a success.

\item  $ex(\Pscr,x=E,\Pscr\uplus \{ \lb x,E' \rb \})\ \leftarrow$  $eval(\Pscr,E,E')$.
\% the assignment statement. Here, 
$\uplus$ denotes a set union but $\lb x,V\rb$ in $\Pscr$ will be replaced by $\lb x,E' \rb$.

\item  $ex(\Pscr,G_1; G_2,\Pscr_2)\ \leftarrow$   $(ex(\Pscr,G_1,\Pscr_1)$  seqand 
  $ex(\Pscr_1,G_2,\Pscr_2))$. \% sequential composition

\item $ex(\Pscr, read(x); G, \Pscr_1)\ \leftarrow$   
  $ex(\Pscr\uplus \{ \lb x,kbd \rb \},G,\Pscr_1)$.   where
$kbd$ is the keyboard input and  $\uplus$
 denotes a set union but $\lb x,V\rb$ in $\Pscr$ will be replaced by $\lb x,kbd \rb$.

\item $ex(\Pscr, \kch(G_1,\ldots,G_n), \Pscr_1)\ \leftarrow$   ((read the keyboard input i)  seqand \\
 $(i \in \{ 1,\ldots,n \}$ seqand $ex(\Pscr, G_i,\Pscr_1))$ choose
$(i \not\in \{ 1,\ldots,n \}$ seqand $(\Pscr_1 == \Pscr)))$

\end{numberedlist}
\end{defn}

\noindent
If $ex(\Pscr,G,\Pscr_1)$ has no derivation, then the machine returns  the failure.

 The rule (8) deals with bounded-choice interaction.    To execute $\kch(G_1,\ldots,G_n)$ successfully, 
the machine does the following:

\begin{numberedlist}

\item It reads and saves the keyboard input value $i$ in some
temporary storage.

\item Then it tries the first branch  of the form $i \in \{ 1,\ldots,n \}$ seqand $ex(\Pscr, G_i,\Pscr_1)$.
 That is, it first
checks whether $i$ is legal, \ie, among $\{ 1,\ldots,  n \}$.  
   The machine then executes $G_i$.

\item If the first branch fails, the machine tries the second branch of the form
$i \not\in \{ 1,\ldots,n \}$ seqand $(\Pscr_1 == \Pscr)$.
That is, it first checks whether $i$ is illegal, \ie, not among $\{ 1,\ldots,  n \}$. 
If it is illegal,  then it means that it is the user, not the machine, who
 failed to do his job. Therefore, the machine sets $\Pscr_1$ to $ \Pscr$ and returns the success.

\end{numberedlist}
\noindent  As an example of our language, the following $G$-formula 

\begin{exmple}
        \kch( \\
 \>           emp = tom;  age = 31, \\
  \>          emp = kim; age = 40, \\
 \>           emp = sue; age = 22 );
\end{exmple}
\noindent expresses the task of the user choosing one among
three employees. More examples are shown in Section 3.

As mentioned earlier, the $\kch$  construct is a well-designed, 
high-level abstraction for bounded-choice interaction which is quite common to
user interaction. As for its implementation,
it can be bolted into the language as a basic statement or it can be supported via preprocessing.
C$^{++}$ macro code for some initial implementation of $\kch$  is available under \\
{\tt http://www.researchgate.net/publication/} \\ 
{\tt 282331184}\footnote{
Unfortunately, C$^{++}$ has little support for variadic macros such as $\kch$. 
For this reason, the current implementation supports only a limited number of arguments (up to 5, to be precise).
 We plan to improve this implementaton in the future.
}.

\section{Examples }\label{sec:modules}

As an  example, consider the following
statement that performs ATM transaction. The types of ATM transaction are
1) balance checking, 2) cash withdrawal, and 3) cash deposition.
An example of this class is provided by the
following code where the program $\Pscr$ is of the form:

\begin{exmple}
deposit() =  \\
 print(``type 1 for \$1 and 2 for \$5:''); \\
\kch(amount = \$1,amount = \$5); $\ldots$  \\

withdraw() = \\
 print(``type 1 for \$1 and 2 for \$5:''); \\
 \kch(amount = \$1,amount = \$5); $\ldots$ \\

balance() =   $\ldots$ \\
\end{exmple}

\noindent and the goal $G$ is of the form:

\begin{exmple}
 print(``type 1 for balance,2 for withdraw,3 for deposit'');\\
       \kch(balance(),    withdraw(), deposit());
\end{exmple}
\noindent In the above, the execution
 basically proceeds as follows: the machine asks the user to choose one among three procedures.
If the user choose the withdrawal  by typing 2, then  the machine will ask the user
again to choose the amount of the withdrawal. Then the execution will go on.
Note that our code is very concise compared to the traditional one.

As a second example, our language  makes it possible to customize the amount for tuition via
interaction with the user.

The following C-like code displays the amount of the tuition, based on the user's
field of study.

\begin{exmple}
        read(major); \\
        switch (major) \{ \\
 \>           case english:  tuition = \$2,000;   break; \\
  \>          case medical:  tuition = \$4,000;   break;\\
 \>           case liberal:  tuition = \$2,200;    break;\\
 \>           default:   tuition  = 0; \} \\
print(tuition);
\end{exmple}

\noindent 
The above code obviously requires a form of bounded-choice interaction rather than unbounded quantification and 
can thus be greatly simplified using the \kch\ statement.
This is shown below:

\begin{exmple}
 print(``type 1 for english,2 for medical,3 for liberal:''); \\
        \kch( \\
 \>           major = english;   tuition = \$2,000, \\
  \>           major= medical;    tuition = \$4,000,  \\
 \>            major = liberal;  tuition = \$2,200);\\
print(tuition); \\
\end{exmple}

\noindent This program expresses the task of the user choosing one among
three majors.
 Note that this program is definitely better than the above: it is concise, much easier to read/write/use,
 and less error-prone.
 The system now requests the user to select one (by typing 1,2 or 3)
among three majors. After it is selected, 
the system displays the amount of the tuition.

\section{Empirical Study}

This section provides some empirical study 
 comparing two languages, namely C and \muprolog. 

 It has    
the following features:   

\begin{itemize}
                                  
\item The same program  is considered for each language.   
 A typical ATM machine in Korea 
   has  a sequence of 3 interactions  for
cash deposit, 4 for cash withdrawal, and 2 for checking balance.
The program we require is an implementation of this ATM machine using seven
major procedures (deposit, withdrawal, balance, password processing, etc).
Overall, there are five occurrences of bounded-choice interactions in the program.

 \item
 For each language, we analyze five best implementations  of the 
program by 
Computer Science undergraduate students
in our Software Engineering classes.

\item Two different aspects are investigated, namely program length and programming effort.

\end{itemize}

{\bf Program length} \\

The following table  shows the numbers of lines of five programs containing 
 a statement, a declaration, or  a delimiter such as a closing brace.

\begin{exmple}
 \>\>          program lines   \>\>\>\>  average line   \\ \\
C   \>\>        (127,130,135,142,154)    \>\>\>\>     137.6 \\
\muprolog   \>\>        (113,115,123,129,132)    \>\>\>\>     122.4 \\
\end{exmple}

    We see that C codes are typically 10\% longer   
than  \muprolog. \\

{\bf Work time and productivity} \\

The following table shows the total work time  for designing, writing, 
and testing the program as  measured by us in the classes.

\begin{exmple}
 \>\>          programming hours  \>\>\>\>  average hour   \\ \\
C   \>\>        (1.6,1.8,2.4,2.5,2.8)    \>\>\>\>     2.2 \\
\muprolog   \>\>        (1.2,1.4,1.5,1.6,2.1)    \>\>\>\>     1.5 \\
\end{exmple}

As we see, \muprolog  
takes less than 70\% as long as C.

\section{Conclusion}\label{sec:conc}

In this paper, we have  extended  core C  by adding a bounded-choice statement.
 This extension allows   $\uch(G_1,\ldots,G_n)$,  where each $G_i$ is a statement.
This statement makes it possible for the core C
to model decision steps from the user. 

 The $\uch(G_1,\ldots,G_n)$ construct allows only a simple form of user input, \ie,   natural numbers.
A more flexible form of user input can be obtained using a parameterized $\uch$ statement
of the form $\uch(c_1:G_1,\ldots,c_n:G_n)$, where $c_1,\ldots,c_n$ are (pairwise disjoint) strings.
The semantics is that if some string $c_i$ is typed, then $G_i$ will be executed.
Thus, the latter allows   the user to type  more symbolic names rather than  just  numbers.
We plan to investigate this possibility in the future.

Although we focused on the keyboard input, it is straightforward to extend our idea
to the mouse input, which plays a central role in smartphone applications.
For example, the statement $mchoose(button_1:G_1,\ldots, button_n:G_n)$ where
each button is a graphic component located at some area can be adopted. The idea is that if $button_i$ is clicked,
then $G_i$ will be executed. It can be easily seen 
 that this statement will greatly
 simplify  smartphone
programming.

We plan to compare our construct to another popular approach: the monad construct 
in functional languages. We also plan to
connect our execution model to Japaridze's elegant 
Computability Logic \cite{Jap03,Jap08},
which has many interesting applications (for example, see \cite{KHP13})
 in information technology. 

\section{Acknowledgements}

We thank the anonymous reviewer for several helpful comments including the parameterized $\uch$ statement.
This work  was supported by Dong-A University Research Fund.

\bibliographystyle{ieicetr}

\begin{thebibliography}{1}

\bibitem{Khan87}
G.~Kahn,  ``Natural Semantics'', In the 4th Annual Symposium on Theoretical Aspects of Computer Science, 
LNCS vol. 247,  1987.



\bibitem{Jap03}
G.~Japaridze, ``Introduction to computability logic'', Annals  of Pure and
 Applied  Logic, vol.123, pp.1--99, 2003.

\bibitem{Jap08}
G.~Japaridze,   ``Sequential operators in computability logic'',
 Information and Computation, vol.206, No.12, pp.1443-1475, 2008.  

\bibitem{KHP13}
K.~Kwon, S.~Hur and M.~Park,  ``Improving Robustness via Disjunctive Statements in Imperative  Programming'', IEICE Transations on Information and Systems, vol.E96-D,No.9, pp.2036-2038, September, 2013.  

\bibitem{HM94}
J.~Hodas and D.~Miller,   ``Logic Programming in a Fragment of Intuitionistic Linear Logic'', 
 Information and Computation, vol.110, No.2, pp.327-365, 1994. 



\bibitem{MNPS91}
D.~Miller, G.~Nadathur, F.~Pfenning, and A.~Scedrov, ``Uniform proofs as a
  foundation for logic programming'', Annals of Pure and Applied Logic, vol.51,
  pp.125--157, 1991.

\bibitem{MN12}
D.~Miller, G.~Nadathur, Programming with higher-order logic, Cambridge University Press,   2012.
\end{thebibliography}



\end{document}